\documentclass[german,english]{bvm2020}

\def\articlenumber{2795}

\date{}
\title{Tenfold your photons---a physically-sound approach to filtering-based variance reduction of Monte-Carlo-simulated dose distributions}

\titlerunning{Tenfold your photons}

\author{Philipp~Roser$^{1,3}$, Annette~Birkhold$^2$, Alexander~Preuhs$^1$, Markus~Kowarschik$^2$, Rebecca~Fahrig$^2$, Andreas~Maier$^{1,3}$}

\authorrunning{Roser et al.}

\institute{%
 $^1$Pattern Recognition Lab, FAU Erlangen-N\"urnberg\\
 $^2$Siemens Healthcare GmbH, Forchheim Germany\\
 $^3$Erlangen Graduate School in Advanced Optical Technologies (SAOT)
}
\email{philipp.roser@fau.de}

\begin{document}

\selectlanguage{english}

\maketitle

\begin{abstract}
X-ray dose constantly gains interest in the interventional suite.
With dose being generally difficult to monitor reliably, fast computational methods are desirable.
A major drawback of the gold standard based on Monte Carlo (MC) methods is its computational complexity.
Besides common variance reduction techniques, filter approaches are often applied to achieve conclusive results within a fraction of time.
Inspired by these methods, we propose a novel approach.
We down-sample the target volume based on the fraction of mass, simulate the imaging situation, and then revert the down-sampling.
To this end, the dose is weighted by the mass energy absorption, up-sampled, and distributed using a guided filter.
Eventually, the weighting is inverted resulting in accurate high resolution dose distributions.
The approach has the potential to considerably speed-up MC simulations since less photons and boundary checks are necessary.
First experiments substantiate these assumptions. 
We achieve a median accuracy of 96.7\,\% to 97.4\,\% of the dose estimation with the proposed method and a down-sampling factor of 8 and 4, respectively.
While maintaining a high accuracy, the proposed method provides for a tenfold speed-up.
The overall findings suggest the conclusion that the proposed method has the potential to allow for further efficiency.
\end{abstract}

\section{Introduction}
Over the last years, X-ray dose awareness increased steadily in the interventional environment -- also driven by legal regulations requiring evidence of consistent dose application through monitoring tools.
Monte Carlo (MC) simulation of particle transport is the de-facto gold standard for computational dose estimation in X-ray imaging.
Only its high algorithmic complexity and demand for extensive prior knowledge about the patient anatomy stands in the way of a wider application in the clinical environment outside radiotherapy, especially in the interventional suite.

While the lack of pre-operative computed tomography (CT) scans in general can be overcome by constantly improving patient surface and organ shape modeling algorithms\,\cite{Zhong:19:AnatomicalLandmarks}, the high arithmetic effort of reliable MC simulations remains a hurdle. 
Although there exists a variety of GPU-accelerated MC codes applicable to X-ray imaging\,\cite{Badal:09:MCGPU} or radiotherapy\,\cite{Bert:13:GGEMS}, their gain in performance mainly depends on the employed hardware, which may vary heavily between different hospitals.
Commonly, different variance reduction techniques such as Russian roulette or delta tracking\,\cite{Woodcock:65:DeltaTracking} are implemented, which, however, might act contrary to the intention of speeding-up the simulation.
For delta tracking e.g., the irradiated volume is assumed to homogeneously consist of the highest-density material during particle tracing to reduce the overall frequency of costly random sampling.
This may lead to an undesired slowdown for very dense materials, commonly found in medical applications, e.g., titanium.
Recently, convolutional neural networks have been introduced to the problem of dose estimation\,\cite{Roser:19:DoseLearning}, however, their dependency on diverse training data renders them infeasible for general purpose dose estimation at this point.

To this end, smoothing approaches such as anisotropic diffusion\,\cite{Miao:03:MCDenoisingAnisotropic} or Savitzky-Golay filtering\,\cite{Kawrakow:02:MCDenoising} have been employed successfully, claiming a further reduction of primary particles by a factor of 2 to 20.
Based on these concepts, we propose a novel theoretical take on image-processing-based variance reduction. 
Before simulation, we apply a physically-sound down-sampling strategy on the target volume combined with super-resolving the resulting dose distribution using guided filtering (GF)\,\cite{He:13:GuidedFilter} and the original target volume as guidance. 
By massively down-sampling the target volume, a further speed-up could possibly be achieved since less boundary checks are necessary.

\section{Methods}
\subsection{Basic principle}
\begin{figure}
    \centering
    \includegraphics[width=\linewidth]{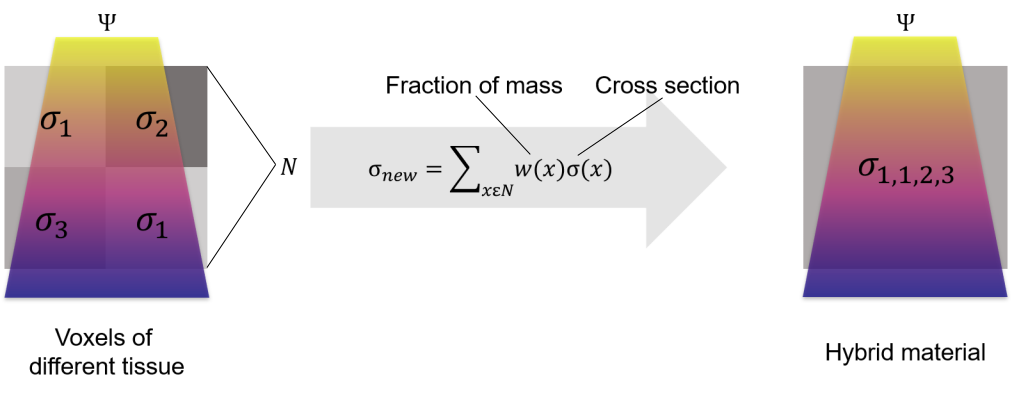}
    \caption{Basic principle of the proposed method: a hybrid material of the neighborhood $\mathcal{N}$ exposes the same macroscopic properties as its finer counterpart.}
    \label{fig:idea}
\end{figure}
The presented idea depicted in Fig.\,\ref{fig:idea} is based on the assumption that the macroscopic properties, such as the photon fluence $\psi(E)$ with respect to the kinetic energy $E$, in a neighborhood $\mathcal{N}$ in the voxelized target volume are approximately equal to when its individual voxels are condensed to a mixture material.
For instance, the differential cross section $\sigma_\mathcal{N}(E)$ of matter in such a neighborhood is defined by
\begin{equation}
    \sigma_\mathcal{N}(E) = \sum_{x \in \mathcal{N}} w(x)\sigma(x,E) \enspace,
\end{equation}
where $w(x)$ is the fraction of mass of voxel $x$, $\sigma(x,E)$ is the differential cross section of its corresponding material, and $E$ is the kinetic energy of the incident particle.
Similarly, the mass energy-absorption coefficient $\left(\frac{\mu_\text{en}(E)}{\rho}\right)_\mathcal{N}$ is defined as
\begin{equation}
    \left(\frac{\mu_\text{en}(E)}{\rho}\right)_\mathcal{N} = \sum_{x \in \mathcal{N}} w(x)\left(\frac{\mu_\text{en}(x,E)}{\rho}\right) \enspace.
\end{equation}
In the following, for the sake of readability, we ignore the energy-dependency in our notation.
Bold-typed quantities refer to 3-D tensors $\in \mathbb{R}^3$.

By calculating mixture models for each neighborhood $\mathcal{N}$ in the target volume $\vec{V}$, we obtain its low resolution representation $\tilde{\vec{V}}$.
Using this down-sampled target volume $\tilde{\vec{V}}$ in a MC simulation, we obtain the low resolution dose distribution $\tilde{\vec{D}} = \text{MC}(\tilde{\vec{V}}) \in \mathbb{R}^3$.

Furthermore, in such large, homogeneous voxels a charged particle equilibrium (CPE) can be assumed. 
Under CPE, the absorbed dose $\tilde{\vec{D}}$ in a volume is approximately equal to the respective collision kerma $\tilde{\vec{K}}_\text{col}$
\begin{equation}
    \tilde{\vec{D}} = \tilde{\vec{K}}_\text{col} + \tilde{\vec{K}}_\text{rad} \enspace, \tilde{\vec{K}}_\text{rad} \xrightarrow{} \vec{0} \enspace,
\end{equation}
given the radiative kerma $\tilde{\vec{K}}_\text{rad}$ approaches $0$, which is the case for diagnostic X-rays.
This allows us to exploit the relationship
\begin{equation}\label{eq:kerma-fluence}
    D_\mathcal{N} = K_{\text{col},\mathcal{N}} = \left(\frac{\mu_\text{en}}{\rho}\right)_\mathcal{N} \psi_\mathcal{N}
\end{equation}
to decouple dose or kerma from the absorbance of the irradiated material.

Subsequently, the low resolution fluence $\tilde{\vec{\psi}}_\mathcal{N}$ is up-sampled to the original resolution $\vec{\psi}_\mathcal{N}$ using nearest neighbor (NN) interpolation and GF
\begin{equation}
    \vec{\psi}_\mathcal{N} = \text{GF}\left(\frac{{\vec{\mu}}_\text{en}}{\vec{\rho}}, \text{NN}\left(\tilde{\vec{\psi}}_\mathcal{N}\right),r\right) \enspace,
\end{equation}
where $\frac{\vec{\mu}_\text{en}}{\rho}$ functions as guidance and $r$ is the filtering radius.
Applying Eq.\,\eqref{eq:kerma-fluence}, we arrive at the high resolution dose distribution
\begin{equation}
    \vec{D} = \frac{{\vec{\mu}}_\text{en}}{\vec{\rho}} \vec{\psi}_\mathcal{N} \enspace.
\end{equation}

\subsection{Proof of concept}
Our MC simulation framework is based on the general purpose MC toolkit Geant4\,\cite{Agostinelli:03:Geant4} due to the great flexibility it offers in terms of particle tracking, experiment geometry, and material modeling.
Unfortunately, the number of different mixture materials increases exponentially with the down-sampling factor, depending on the degree of distinction of different organs and tissues in the target volume $\vec{V}$.
This in turn leads to the fact that the calculation of the mixture materials cannot be carried out without further development.

To still provide for a proof of concept, we synthetically create corresponding low resolution dose distributions $\tilde{\vec{D}_s}$ from high resolution dose distributions $\vec{D}$, where $s$ is the sampling factor.
The down-sampling is performed by weighting and summing all voxels in the neighborhood $\mathcal{N}(s)$.
Again, the weights correspond to the fraction of mass of each individual voxel of the resulting voxel.

\section{Results}
\begin{figure}
    \centering
    \includegraphics[width=\textwidth]{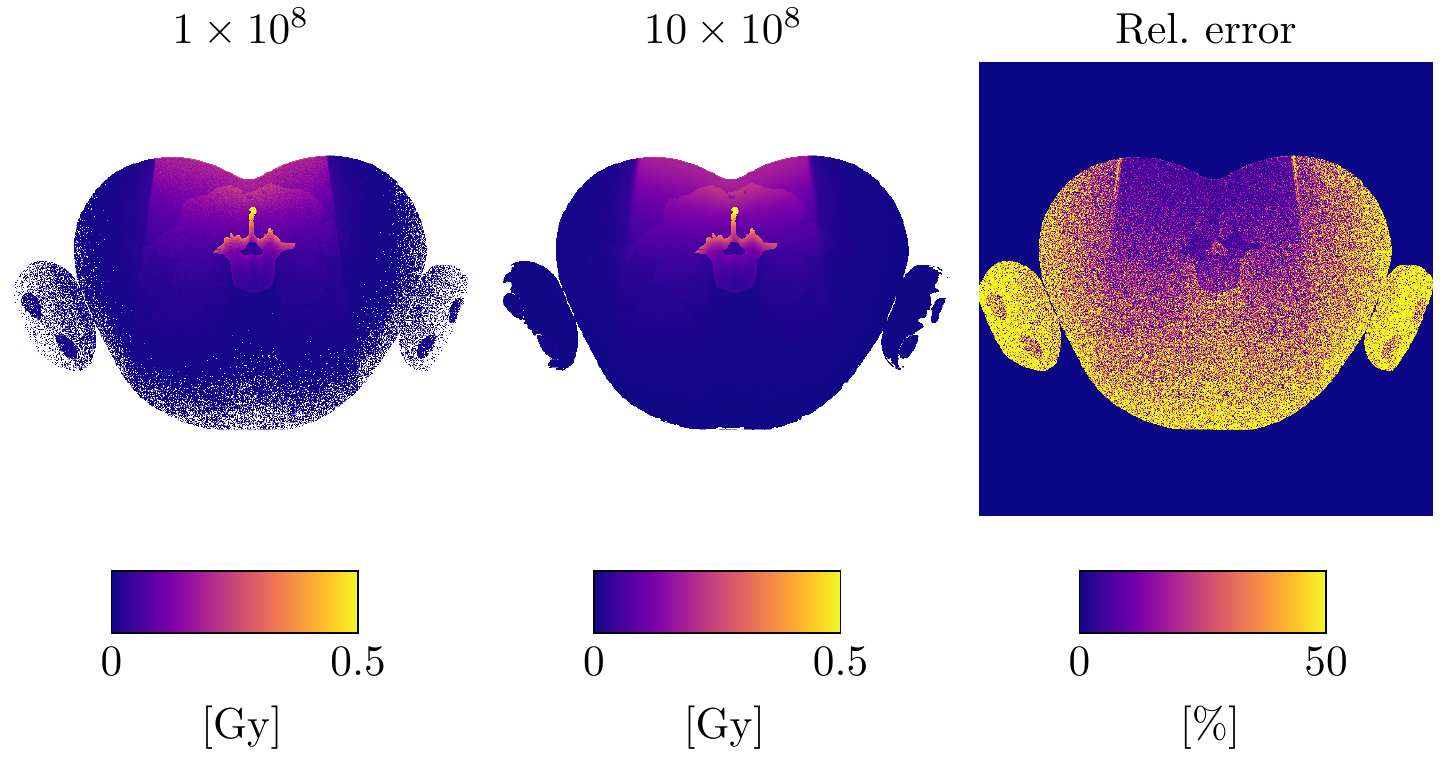}
    \caption{Baseline error between average dose distributions of $10^8$ and $10\times 10^8$ without any processing. Both distributions are scaled to a peak dose of 0.5\,Gy.}
    \label{fig:baseline}
\end{figure}
\begin{figure}
    \centering
    \includegraphics[width=\textwidth]{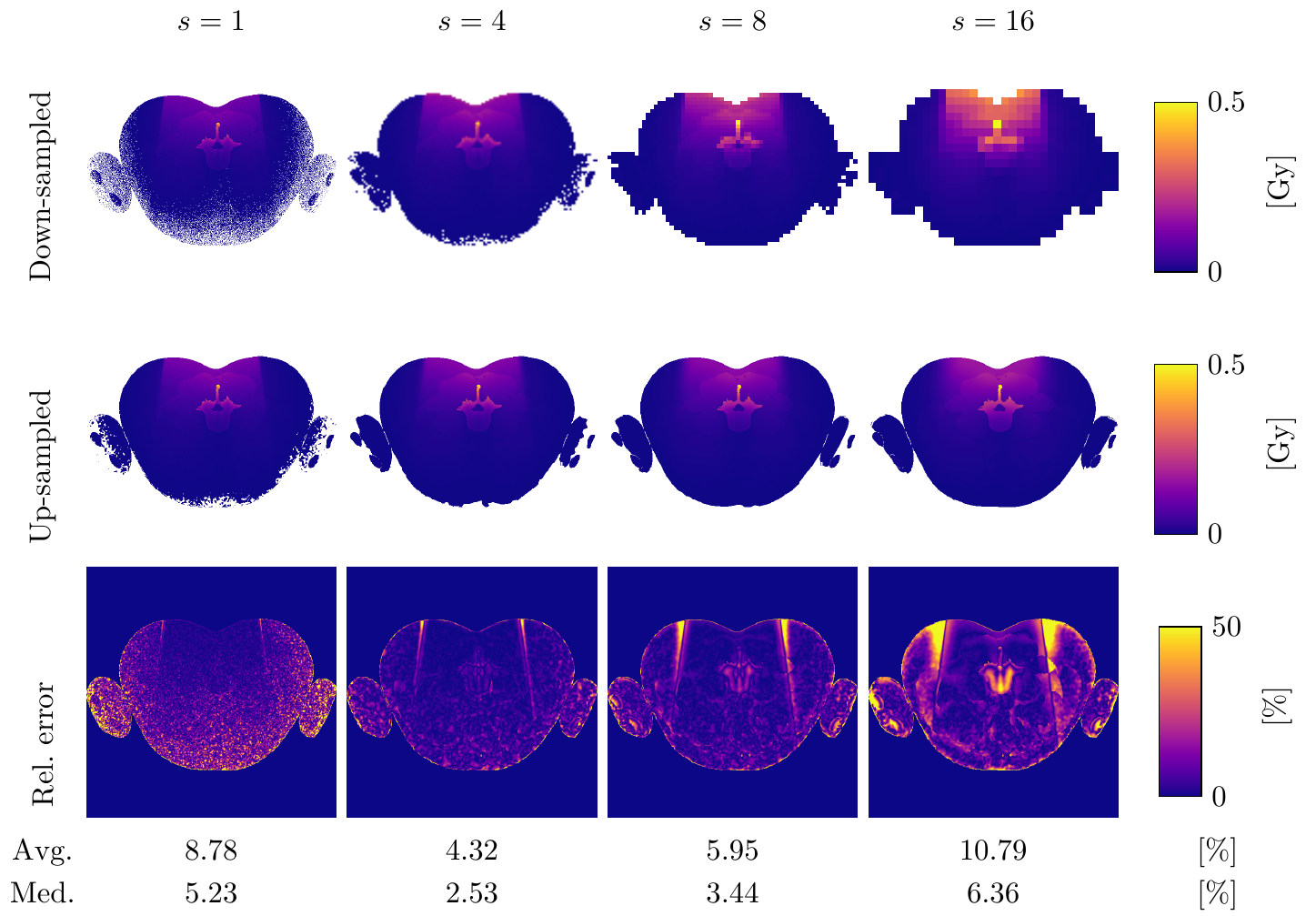}
    \caption{Exemplary axial slices of dose distributions down-sampled by the factor $s$ (first row) and corresponding reconstructed high resolution distributions (second row).
    The last row shows the relative error with respect to the reference dose distribution and corresponding averages and medians. Dose distributions are scaled to a peak dose of 0.5\,Gy.}
    \label{fig:results}
\end{figure}
We investigate our method at four different scales $s \in \{1, 4, 8, 16\}$ of a dose distribution simulated with respect to $10^8$ primary photons sampled from a 120\,kV peak voltage spectrum using the digital Visible Human dosimetry phantom\,\cite{Zankl:02:Vishum}.
As reference, we consider a simulation with $10\times 10^8$ primary photons and otherwise same parameters.
Figure\,\ref{fig:baseline} shows the initial deviation between these two simulations.
Note that both distributions are normalized to a peak dose of 0.5\,Gy.
We observe an average and median relative error of 34.8\,\% and 22.3\,\%, respectively, when no further processing is applied.

In comparison, Fig.\,\ref{fig:results} shows the down-sampled and up-sampled dose distributions using our method, and respective error rates.
For the GF operation we set $r = s$.
We can see that even for $s = 16$ high resolution dose distributions can be reconstructed with 10.79\,\% average and 6.36\,\% median error only.
As to be expected, with decreasing sampling factors $s \in \{4, 8\}$, these error rates drop to 4.32\,\% average and 2.53\,\% median.
Surprisingly, these errors are significantly lower than those arising from smoothing the original dose distribution ($s = 1$) using GF, where no low resolution needs to be compensated.
Overall, the highest errors can be observed at the transition of primary X-ray beam to scattered radiation, due to the diffuse border in the low resolution dose distributions.

\section{Discussion}
We proposed a theoretical framework for accelerated MC simulation based on a down- and up-sampling scheme for the target volume and resulting dose distributions.
Since, its implementation is currently not feasible in our MC application, we gave a proof of concept by transferring the basic principle of our method to synthetically down-sampled dose distributions.
Promising results could be reported, which substantiate the assumption of the method being applicable to speed up MC simulations considerably.
Future studies will have to focus on a feasible implementation in Geant4 as well as an in-depth analysis of the expected gain in computational performance.

The presented results also suggest the conclusion that the overall method can be used to de-noise MC simulations in general.
The down- and up-sampling of the dose distribution could be reformulated to filtering operation.

Inspecting the results visually, it becomes however evident that our method exposes weaknesses at edges and interfaces of different tissues.
In addition, for higher down-sampling factors, a systematic error trend in higher density tissues such as bone is observable.
These issues could be solved by formulating the GF radius $r$ as function of the tissue densities in the neighborhood $\mathcal{N}$.
Furthermore, the inclusion of a voxel-wise distance weighting with respect to the radiation source could be beneficial when applying GF.

~\\
{\bf Disclaimer:} The concepts and information presented in this paper are based on research and are not commercially available.

\bibliographystyle{bvm2020}

\bibliography{2795}

\begin{thebibliography}{10}

\bibitem{Zhong:19:AnatomicalLandmarks}
Zhong X, Strobel N, Birkhold A, Kowarschik M, Fahrig R, Maier A.
\newblock A machine learning pipeline for internal anatomical landmark
  embedding based on a patient surface model.
\newblock Int J Comput Assist Radiol Surg. 2019;14(1):53--61.

\bibitem{Badal:09:MCGPU}
Badal A, Badano A.
\newblock Accelerating {Monte Carlo} simulations of photon transport in a
  voxelized geometry using a massively parallel graphics processing unit.
\newblock Med Phys. 2009;36(11):4878--4880.

\bibitem{Bert:13:GGEMS}
Bert J, Perez-Ponce H, Bitar ZE, Jan S, Boursier Y, Vintache D, et~al.
\newblock Geant4-based {Monte Carlo} simulations on {GPU} for medical
  applications.
\newblock Phys Med Biol. 2013;58(16):5593--5611.

\bibitem{Woodcock:65:DeltaTracking}
Woodcock E, Murphy T, Hemmings P, Longworth T.
\newblock Techniques used in the GEM code for Monte Carlo neutronics
  calculations in reactors and other systems of complex geometry, ANL-7050.
\newblock Argonne National Laboratory; 1965.

\bibitem{Roser:19:DoseLearning}
Roser P, Zhong X, Birkhold A, Strobel N, Kowarschik M, Fahrig R, et~al.
\newblock Physics-driven learning of x-ray skin dose distribution in
  interventional procedures.
\newblock Med Phys. 2019;46(10):4654--4665.

\bibitem{Miao:03:MCDenoisingAnisotropic}
Miao B, Jeraj R, Bao S, Mackie TR.
\newblock Adaptive anisotropic diffusion filtering of Monte Carlo dose
  distributions.
\newblock Phys in Med and Biol. 2003;48(17):2767--2781.

\bibitem{Kawrakow:02:MCDenoising}
Kawrakow I.
\newblock On the de-noising of Monte Carlo calculated dose distributions.
\newblock Phys Med and Biol. 2002;47(17):3087--3103.

\bibitem{He:13:GuidedFilter}
{He} K, {Sun} J, {Tang} X.
\newblock Guided Image Filtering.
\newblock IEEE Trans Pattern Anal Mach Intell. 2013;35(6):1397--1409.

\bibitem{Agostinelli:03:Geant4}
Agostinelli S, Allison J, Amako K, Apostolakis J, Araujo H, Arce P, et~al.
\newblock Geant4--a simulation toolkit.
\newblock Nucl Instrum Meth A. 2003;506(3):250--303.

\bibitem{Zankl:02:Vishum}
Zankl M, Petoussi-Henss N, Fill U, Regulla D.
\newblock Tomographic anthropomorphic models part {IV}: Organ doses for adults
  due to idealized external photon exposures.
\newblock Institute of Radiation Medicine (former Institute of Radiation
  Protection); 2002.

\end{thebibliography}
\marginpar{\color{white}E\articlenumber} 
\end{document}